



\font\bigbold=cmbx12
\font\eightrm=cmr8
\font\sixrm=cmr6
\font\fiverm=cmr5
\font\eightbf=cmbx8
\font\sixbf=cmbx6
\font\fivebf=cmbx5
\font\eighti=cmmi8  \skewchar\eighti='177
\font\sixi=cmmi6    \skewchar\sixi='177
\font\fivei=cmmi5
\font\eightsy=cmsy8 \skewchar\eightsy='60
\font\sixsy=cmsy6   \skewchar\sixsy='60
\font\fivesy=cmsy5
\font\eightit=cmti8
\font\eightsl=cmsl8
\font\eighttt=cmtt8
\font\tenfrak=eufm10
\font\eightfrak=eufm8
\font\sevenfrak=eufm10 at 7pt
\font\fivefrak=eufm5
\font\tenbb=msbm10
\font\eightbb=msbm8
\font\sevenbb=msbm7
\font\fivebb=msbm5
\font\tensmc=cmcsc10


\newfam\bbfam
\textfont\bbfam=\tenbb
\scriptfont\bbfam=\sevenbb
\scriptscriptfont\bbfam=\fivebb
\def\Bbb{\fam\bbfam}

\newfam\frakfam
\textfont\frakfam=\tenfrak
\scriptfont\frakfam=\sevenfrak
\scriptscriptfont\frakfam=\fivefrak
\def\frak{\fam\frakfam}

\def\smc{\tensmc}


\def\eightpoint{%
\textfont0=\eightrm   \scriptfont0=\sixrm
\scriptscriptfont0=\fiverm  \def\rm{\fam0\eightrm}%
\textfont1=\eighti   \scriptfont1=\sixi
\scriptscriptfont1=\fivei  \def\oldstyle{\fam1\eighti}%
\textfont2=\eightsy   \scriptfont2=\sixsy
\scriptscriptfont2=\fivesy
\textfont\itfam=\eightit  \def\it{\fam\itfam\eightit}%
\textfont\slfam=\eightsl  \def\sl{\fam\slfam\eightsl}%
\textfont\ttfam=\eighttt  \def\tt{\fam\ttfam\eighttt}%
\textfont\frakfam=\eightfrak \def\frak{\fam\frakfam\eightfrak}%
\textfont\bbfam=\eightbb  \def\Bbb{\fam\bbfam\eightbb}%
\textfont\bffam=\eightbf   \scriptfont\bffam=\sixbf
\scriptscriptfont\bffam=\fivebf  \def\bf{\fam\bffam\eightbf}%
\abovedisplayskip=9pt plus 2pt minus 6pt
\belowdisplayskip=\abovedisplayskip
\abovedisplayshortskip=0pt plus 2pt
\belowdisplayshortskip=5pt plus2pt minus 3pt
\smallskipamount=2pt plus 1pt minus 1pt
\medskipamount=4pt plus 2pt minus 2pt
\bigskipamount=9pt plus4pt minus 4pt
\setbox\strutbox=\hbox{\vrule height 7pt depth 2pt width 0pt}%
\normalbaselineskip=9pt \normalbaselines
\rm}


\def\pagewidth#1{\hsize= #1}
\def\pageheight#1{\vsize= #1}
\def\hcorrection#1{\advance\hoffset by #1}
\def\vcorrection#1{\advance\voffset by #1}

\newif\iftitlepage   \titlepagetrue               
\newtoks\titlepagefoot     \titlepagefoot={\hfil} 
\newtoks\otherpagesfoot    \otherpagesfoot={\hfil\tenrm\folio\hfil}
\footline={\iftitlepage\the\titlepagefoot\global\titlepagefalse
           \else\the\otherpagesfoot\fi}

\def\abstract#1{{\parindent=30pt\narrower\noindent\eightpoint\openup
2pt #1\par}}


\newcount\notenumber\notenumber=1
\def\note#1
{\unskip\footnote{$^{\the\notenumber}$}
{\eightpoint\openup 1pt #1}
\global\advance\notenumber by 1}


\def\frac#1#2{{#1\over#2}}

\def\tfrac#1#2{{\textstyle{#1\over#2}}}
\def\({\left(}
\def\){\right)}
\def\<{\langle}
\def\>{\rangle}

\def\pmb#1{\setbox0=\hbox{$#1$}%
   \kern-.025em\copy0\kern-\wd0
   \kern.05em\copy0\kern-\wd0
   \kern-.025em\raise.0433em\box0 }


\global\newcount\secno \global\secno=0
\global\newcount\meqno \global\meqno=1
\global\newcount\appno \global\appno=0
\newwrite\eqmac
\def\romappno{\ifcase\appno\or A\or B\or C\or D\or E\or F\or G\or H
\or I\or J\or K\or L\or M\or N\or O\or P\or Q\or R\or S\or T\or U\or
V\or W\or X\or Y\or Z\fi}
\def\eqn#1{
        \ifnum\secno>0
            \eqno(\the\secno.\the\meqno)\xdef#1{\the\secno.\the\meqno}
          \else\ifnum\appno>0

\eqno({\rm\romappno}.\the\meqno)\xdef#1{{\rm\romappno}.\the\meqno}
          \else
            \eqno(\the\meqno)\xdef#1{\the\meqno}
          \fi
        \fi
\global\advance\meqno by1 }


\global\newcount\refno
\global\refno=1 \newwrite\reffile
\newwrite\refmac
\newlinechar=`\^^J
\def\ref#1#2{\the\refno\nref#1{#2}}
\def\nref#1#2{\xdef#1{\the\refno}
\ifnum\refno=1\immediate\openout\reffile=refs.tmp\fi
\immediate\write\reffile{
     \noexpand\item{[\noexpand#1]\ }#2\noexpand\nobreak.}
     \immediate\write\refmac{\def\noexpand#1{\the\refno}}
   \global\advance\refno by1}
\def\semi{;\hfil\noexpand\break ^^J}
\def\nl{\hfil\noexpand\break ^^J}
\def\refn#1#2{\nref#1{#2}}

\def\jmp#1#2#3{{\it J. Math. Phys.} {\bf {#1}} (19{#2}) #3}
\def\ijmp#1#2#3{{\it Int. J. Mod. Phys.} {\bf A{#1}} (19{#2}) #3}

\def\pl#1#2#3{{\it Phys. Lett.} {\bf {#1}B} (19{#2}) #3}
\def\np#1#2#3{{\it Nucl. Phys.} {\bf B{#1}} (19{#2}) #3}

\def\prD#1#2#3{{\it Phys. Rev.} {\bf D{#1}} (19{#2}) #3}

\def\rmp#1#2#3{{\it Rev. Mod. Phys.} {\bf {#1}} (19{#2}) #3}
\def\ann#1#2#3{{\it Ann. Phys.} {\bf {#1}} (19{#2}) #3}
\def\prp#1#2#3{{\it Phys. Rep.} {\bf {#1}C} (19{#2}) #3}
\def\prog#1#2#3{{\it Prog. Theor. Phys.} {\bf {#1}} (19{#2}) #3}
\def\jpa#1#2#3{{\it J. Phys. A:} {\bf {#1}} (19{#2}) #3}


\pageheight{23cm}
\pagewidth{15.5cm}
\hcorrection{-2.5mm}
\magnification \magstep1
\baselineskip=16pt plus 1pt minus 1pt
\parskip=5pt plus 1pt minus 1pt


\def\g{{\frak g}}
\def\h{{\frak h}}
\def\r{{\frak r}}
\def\t{{\frak t}}
\def\Real{{\Bbb R}}

\def\H{{\cal H}}
\def\a{\alpha}
\def\b{\beta}
\def\d{\delta}
\def\e{\varepsilon}
\def\s{\sigma}
\def\sa{\s_\a}
\def\sb{\s_\b}
\def\pa{\partial}
\def\ket#1{|#1\rangle}
\def\bra#1{\langle#1|}

%
%
{
\eightpoint

\refn\Shapere
{See, for example,
A. Shapere and F. Wilczek (eds), {\it Geometric Phases in Physics\/}
(Singapore,
World Scientific 1989)
}

\refn\LL
{N.P. Landsman and N. Linden, \np{365}{91}{121}}

\refn\Vinet
{L. Vinet, \prD{37}{88}{2369}}

\refn\Giler
{S. Giler, P. Kosinski and L. Szymanowsky, \ijmp{4}{89}{1453}}

\refn\Levay
{P. L\'evay, \jpa{27}{94}{2857}}

\refn\Mackey
{G.W. Mackey, {\it Induced Representations of Groups
and Quantum Mechanics} (Benjamin, New York, 1969)}

\refn\OK
{Y. Ohnuki and S. Kitakado, \jmp{34}{93}{2827}}

\refn\TT
{S. Takagi and T. Tanzawa, \prog{87}{92}{561}}

\refn\FO
{K. Fujii and N. Ogawa, \prog{89}{93}{575}}

\refn\MT
{D. McMullan and I. Tsutsui, \ann{237}{95}{269}; \pl{320}{94}{287}}

\refn\KN
{S. Kobayashi and K. Nomizu, {\it Foundations of Differential
Geometry
Vol.I\/} (Interscience, New York, 1969)}

\refn\Camporesi
{R. Camporesi, \prp{196}{90}{1}}

\refn\Strathdee
{J. Strathdee, in  {\it Supersymmetry and Supergravity `82\/},
S. Ferrara {\it et.al.} (eds), (World Scientific, Singapore,
1983)}

\refn\Tanimura
{S. Tanimura, \lq\lq Quantum Mechanics on Manifolds\rq\rq,
Nagoya Univ. preprint DPNU-93-21}

\refn\FKO
{K. Fujii, S. Kitakado and Y. Ohnuki,
{\lq\lq Gauge Structure on $S^D$\rq\rq},
{\it Int. Journ. Mod. Phys.} (1995), to appear}

\refn\Fujii
{K. Fujii, {\it Lett. in Math. Phys.} {\bf 12} (1986) 363}

\refn\Bohm
{A. Bohm, {\it Quantum Mechanics: Foundations and Applications\/},
second
edition
(Springer-Verlag, New York, 1986)}

\refn\Wudka
{J. Wudka, \jpa{25}{92}{2945}}

\refn\Hum
{S.E. Humphreys, {\it Introduction to Lie Algebras
and Representation Theory}
(Springer-Verlag, Berlin, 1972)}

\refn\Wallach
{N.R. Wallach, {\it Harmonic Analysis on Homogeneous Spaces\/}
(Marcel Dekker Inc, New York, 1973)}

\refn\JK
{H. Jensen and H. Koppe, \ann{63}{71}{586}}

\refn\ZG
{W. Zhang, D. H. Feng and R. Gilmore, \rmp{62}{90}{867}}

}

%
%


\null
{\baselineskip=10pt
\rightline{PLY-MS-95-02}
\rightline{INS-Rep.-1094}
\rightline{April 1995}
\vfill
}
\centerline{\bigbold The Canonical Connection in Quantum Mechanics}
\vskip 30pt
\centerline{\smc P\'eter L\'evay}
\vskip 5pt
{\baselineskip=12pt
\centerline{Quantum Theory Group}
\centerline{Institute of Physics}
\centerline{Technical University of Budapest}
\centerline{H-1521 Budapest}
\centerline{Hungary}
\centerline{(e-mail: levay@phy.bme.hu)}}

\vskip 15pt
\centerline{\smc David McMullan}
\vskip 5pt
{\baselineskip=12pt
\centerline{School of Mathematics and Statistics}
\centerline{University of Plymouth}
\centerline{Drake Circus, Plymouth, Devon PL4 8AA}
\centerline{U.K.}
\centerline{(e-mail: d.mcmullan@plymouth.ac.uk)}}
\vskip 10pt
\centerline
{\smc and}

\vskip 8pt
\centerline{\smc Izumi Tsutsui}
\vskip 5pt
{\baselineskip=12pt
\centerline{Institute for Nuclear Study}
\centerline{University of Tokyo}
\centerline{Midori-cho, Tanashi-shi, Tokyo 188}
\centerline{Japan}
\centerline{(e-mail: tsutsui@ins.u-tokyo.ac.jp)}
}
\vskip 30pt
\abstract{%
{\bf Abstract.}\quad
In this paper we investigate
the form of induced gauge fields that
arises in two types of quantum systems.
In the first we consider
quantum mechanics on  coset spaces G/H, and argue that
G-invariance is central to the
emergence of the H-connection as induced gauge fields
in the different quantum sectors.
We then demonstrate why the same connection,
now giving rise to the non-abelian generalization of Berry's phase,
can also be found
in systems which have slow variables
taking values in such a coset space.
}

\vfill\eject


\noindent \secno=1 \meqno=1

\noindent
{\bf 1. Introduction}
\medskip
\noindent
There are various instances in quantum mechanics
when a gauge field appears in a system whose initial formulation
did not contain such fields. The most familiar example of this is the
emergence of Berry's connection [\Shapere] in
systems with degeneracies,
which leads to a holonomy in energy
eigenspaces, {\it i.e.}, a non-abelian generalization of Berry's phase.
Another example is to be found in the different quantum sectors
that arise when quantizing on a coset space [\LL].
For both of these cases, the gauge field that
emerges is often found to be of a specific type. Indeed,
when the effective configuration space is a coset space G/H, the
resulting connection can usually be identified with the so-called
{\it {\rm H}-connection}, which is a (possibly topological)
solution of the Yang-Mills equation
on this space.
The prime aim of this paper is to clarify why and when this connection
arises in these systems.

More precisely, in the context of Berry's phase,
the origin of the connection is in some sense obvious from the
outset, that is, it comes from the ambiguity
in choosing a set of basis vectors in the instantaneous energy
eigenspaces.
However,
what is not obvious and hence
remarkable is that in a wide variety of systems of physical
interest Berry's connection often (though not always) takes the form
of the H-connection [\Vinet, \Giler, \Levay].
Such systems arise when considering the coupled dynamics of
slow and fast variables.
In this case we wish to know the form of the connection,
also occurring in the Hamiltonian of the effective slow system, in
advance.
By giving a precise identification of when
it is the H-connection the need to calculate energy
eigenstates can be avoided.

In constrast, in the context of inequivalent quatizations on coset
spaces, the origin of the connection is not quite obvious, and
the question is why the specific H-connection can appear
at all when quantized.
In the account presented in [\LL]
which relies (basically) on Mackey's approach [\Mackey],
the system of \lq free particle' on G/H is considered, where
the Hamiltonian is fixed by
requiring that
there is no operator ordering ambiguity.
This is clearly an important
criterion, and leads to a system minimally coupled to the H-
connection.
However, this is not a criterion geometrically motivated, and
more importantly,
in any attempt at extending these results to field theories
such a reliance on a factor ordering argument is unnatural and,
indeed,
unworkable.
What we will show in this paper is that an invariance
argument can be developed which highlights the need for such a
connection.

The emergence of gauge fields is also recognized recently
by a number of other groups [\OK, \TT, \FO]
using different approaches to quantization.
For instance, in [\OK] spheres $S^n$ embedded in $\Real^{n+1}$
are taken as the configuration
space and
gauge fields are seen to  emerge at the quantum level.
It will be shown, however, in this paper that these induced gauge fields
are none other than the H-connection.  This will
perhaps support the view that the emergence of gauge fields is not
just an artifact of a particular quantization approach but a \lq norm'
when quantizing on coset spaces.

The plan of this paper is as follows.  In Section 2 we will
demonstrate
how the H-connection emerges in the quantum description of a point
particle moving freely on a coset space.  In Section 3 we prove that
the connection
that arises in the quantization scheme of
[\OK] is just the H-connection.  In Section 4
the conditions under which Berry's connection
reduces to the H-connection will
be presented.  Section 5 is devoted to our conclusions and
discussions.

\vfil\eject


\noindent \secno=2 \meqno=1

\noindent
{\bf 2. Quantizing on a coset}
\medskip
\noindent
We begin by arguing that the H-connection ---
observed
by Landsman and Linden [\LL] in investigating the
dynamical aspect of the quantum theory
on a coset space G/H --- is
indeed the natural connection in the quantum system.

Let us first, though, fix our notation (which follows those in [\MT]).
We take G to be a compact Lie group with Lie algebra $\g$, and
H a compact subgroup of G with Lie algebra $\h$.
The Lie algebra $\g$ has an orthogonal decomposition,
$$
\g=\h\oplus\r\,,
\eqn\od
$$
where $\r=\h^\perp$ is the orthogonal complement of $\h$ in $\g$.
This is, in fact, a reductive decomposition, {\it i.e.,}
$$
[\h,\r]\subset\r\,.
\eqn\red
$$
We shall denote bases of the spaces by
$$
\eqalign{
\g &= \hbox{span} \{T_m\}\,, \cr
\h &= \hbox{span} \{T_i\}\,, \cr
\r &= \hbox{span} \{T_a\}\,, \cr
}
\qquad
\eqalign{
m &= 1, \ldots, \dim{\rm G}\,, \cr
i &= 1, \ldots, \dim{\rm H}\,, \cr
a &= 1, \ldots, \dim{\rm (G/H)}\,. \cr
}
\eqn\conv
$$

Let us recall that in Mackey's account of quantizing on G/H [\Mackey]
a set of fundamental relations, called a {\it system
of imprimitivity},
is introduced whose irreducible representations give the
quantum theories (a full discussion of this can be found in [\MT]).
The
upshot of this is that
the Hilbert space $\H{\rm(G/H)}$ on the coset space
consists of $L^2$-functions on G/H belonging to the linear space
$\H_\chi$ of some irreducible unitary
representation $\chi$ of the subgroup H:
$\H({\rm G/H}) \simeq L^2({\rm G/H},\H_\chi)$.
Locally, we may take a basis set $\{\ket{q,\chi,\mu}\}$
of the Hilbert space $\H({\rm G/H})$
by
$$
\ket{q, \chi, \mu} := \ket{q} \otimes \ket{\chi, \mu},
\eqn\basis
$$
where $\ket{q}$ are the eigenstates in the coordinate
representation
on G/H and $\ket{\chi, \mu}$ the orthonormal basis vectors in
$\H_\chi$.
Thus the states in the basis set (\basis) satisfy the orthonormality
condition
$$
\bra{q, \chi, \mu} q', \chi, \nu \> = \delta_{\mu\nu}\delta(q-q'),
\eqn\orthonormal
$$
with $\delta(q-q')$ being the delta-function on the coset space G/H.

In order to have a singularity-free description we need to
introduce a set of patches
to cover the coset space G/H.  Let
$\{ U_\alpha\}$ be the local patches introduced, and
$\sa:U_\a \mapsto {\rm G}$ be a continuous section on the
patch $U_\a$.  On overlaps
$U_\a\cap U_\b$ the sections are related by a gauge transformation,
namely, for $q\in U_\a\cap U_\b$,
$$
\sb(q)=\sa(q) h_{\a\b}(q),
\eqn\gtrp
$$
where $h_{\a\b}\in{\rm H}$.
Accordingly, we consider a sectional basis
$\{ \vert q, \chi, \mu \>^\alpha \}$ which is a basis set
given independenly on the patch $U_\a$.
Using standard partition of unity arguments, we can define an
innerproduct on
these and see that all is well defined.
The wave functions are then defined to be
$$
\psi_\mu^\alpha (q) = {}^\alpha\!\langle{q,\chi,\mu}\ket{\psi}\,.
\eqn\vvwf
$$

An important ingredient in Mackey's quantization [\Mackey] is that
associated with the G-action $q \rightarrow g^{-1}q$
for $g \in {\rm G}$, which relates any
two points on the coset space,
there is a corresponding action on the wave functions
furnished by the {\it induced representation},
$$
\bigl(U(g)\psi\bigr)^\a_\mu(q)
   = \sum_\nu \pi^\chi_{\mu\nu}\bigl((\sa(q))^{-1}g\sb(g^{-1}q)\bigr)\
     \psi^\b_\nu(g^{-1}q)\,.
\eqn\induced
$$
Here the matrix elements of the unitary operator $\pi^\chi(h)$,
implementing the irreducible representation $\chi$, are
$$
\pi^\chi_{\mu\nu}(h):=\bra{\chi,\mu}\pi^\chi(h)\ket{\chi,\nu}\,,
\eqn\matel
$$
and a choice of section has been made on each of the patch,
$q \in U_\a$ and $g^{-1}q \in U_\b$.
On the sectional basis, this action (\induced) reads
$$
U(g)\ket{q, \chi, \mu}^\a = \sum_\nu
   \ket{gq, \chi, \nu}^\b\,
  \pi^\chi_{\nu\mu}\bigl((\sb(gq))^{-1}g\sa(q)\bigr)\ ,
\eqn\indrep
$$
where we put $g \rightarrow g^{-1}$ for later convenience.
In effect, the induced representation (\indrep)
consists of a rotation in the
space $\H_\chi$ and a translation in the coset space G/H, both
determined by $g$ and $q$.
Using the naturally defined measure on the coset space G/H,
one can readily show that (\indrep) indeed provides a
unitary representation of G [\MT].

Now we shall consider the quantum mechanics of a point particle
moving freely on the coset space G/H.  Here the term
\lq free' is meant to indicate that
the system under consideration is {\it homogeneous} over G/H, and that
the dynamics of the particle is that of a free particle when
observed locally.
Note that in order to get
the Schr\"odinger equation for the wave functions
(\vvwf) describing the point particle of this system,
we need to use the G-action to ensure the homogeneity (as it is
the only means available on G/H for this purpose).
But since the G-action on the wave functions
(\induced) is section dependent, we need a covariant derivative
(with respect to $q$)
such that the section dependence disappears in the physical dynamics.

To be explicit, let us consider the state
$$
\ket{\chi, \bar\mu}
  :=\sum_\nu \ket{e, \chi, \nu}^\b\,
  \pi_{\nu\mu}\bigl((\sb(e))^{-1}\bigr)\ ,
\eqn\inq
$$
where $e$ is the identity point in the coset
G/H.  Then (\indrep) allows us
to write the basis states at $q$ as
$$
\ket{q, \chi, \mu}^\a = U\bigl(\sa(q)\bigr) \ket{\chi, \bar\mu}\ ,
\eqn\sol
$$
which shows that the G-action allows for
obtaining all the basis states over G/H by the unitary G-action
from the reference state (\inq).
It is then easy to see from (\indrep)
that, under the change of section (\gtrp),
the basis states undergo the rotation,
$$
\ket{q, \chi, \mu}^\a \rightarrow
\ket{q, \chi, \mu}^\b
= U(\sa h_{\a\b})  \ket{\chi, \bar\mu}
= \sum_\nu \ket{q, \chi, \nu}^\a \pi^\chi_{\nu\mu}(h_{\a\b})\ .
\eqn\haction
$$
Thus, the connection used in the
covariant derivative must compensate the derivative factor
in the Schr\"odinger equation
arising from the rotation in (\haction).
Actually,
in the theory of vector bundles
associated with the
principal bundle G(G/H, H), the term \lq connection' already
implies this property.
This, however, is not enough to single out the
connection relevant to our system on G/H.

The crucial point in specifying the connection is the homogeneity
over the coset G/H mentioned above.
We note that for the system to be homogeneous
the connection must also be homogeneous
physically, that is,
it must be invariant under the
G-action up to a gauge transformation of the group H
({\it i.e.,} up to a change of section).
In other words, the curvature of the connection is constant over G/H.
Now the theory of invariant
connections (see Theorem 11.1 on p.103 of Ref.[\KN])
asserts that such a connection is always given
by the H-connection $A^{\rm H} :=
\sa^{-1}(q) d \sa(q)\vert_\h$, which
is the (pullback of the)
canonical 1-form projected down to the subspace $\h \subset \g$.
In the present context, the invariant
connection that arises in the covariant derivative
acting on the wave functions (\vvwf) is
the H-connection in the representation $\chi$:
$$
\sum_i A^{\rm H}_i(q) (T_i)_{\mu\nu} = \< \chi, \bar\mu \vert \, \,
U^{-1}\bigl((\sa(q)\bigr)\, d U\bigl(\sa(q)\bigr)\vert_\h \,
    \vert \chi, \bar\nu \> \ .
\eqn\hcon
$$
One can readily confirm that its curvature is indeed
constant over G/H and that it does transform as
a connection under the change of section (\gtrp).

In short, we see that the covariant derivative
used for the Schr\"odinger equation must
contain the H-connection in the form
(\hcon), if we are to consider the homogeneous free
particle system over G/H requiring
the independence of the choice of section.
This G-invariance is, we feel, more fundamental than
the factor ordering criterion
adopted in [\LL]. However, for completeness, we now need
to see what form of Hamiltonian comes out of our analysis.

To begin with, let us note that our
vector-valued wave functions
${\psi}^\alpha_\mu(q)$, provided by the irreducible representation
$\chi$ of H
,
may be expanded in terms of the \lq harmonics'
$U^\Lambda_\xi\bigl({\sigma}_\a^{-1}(q)\bigr)$ over the coset
space G/H [\Camporesi],
$$
{\psi}^\alpha_\mu(q)= \sum_\Lambda
\sum_{\rho,\xi} c^{\Lambda}_{\rho\xi}\,
U^\Lambda_\xi\bigl({\sigma}_\a^{-1}(q)\bigr)_{\mu\rho}\ .
\eqn\harwave
$$
In this expansion $\xi$ is the index of multiplicity
of the representation $\chi$ appearing in the irreducible
representation $\Lambda$ of G upon restriction to H, and
the range of $\rho$ equals the dimension of the representation
$\Lambda$.

We recall that
the Frobenius reciprocity theorem tells that in the above
summation only those $\Lambda$ of G occur which
contain the representation $\chi$ of H when restricted to the
subgroup.
(For brevity we henceforth omit $\a$ which
labels the patch to which the point $q$ belongs.)
But the message important to us here is that we can now work with
the section variable $\s^{-1}(q)$
instead of the cordinates $q$ on the coset space.
We shall for the sake of simplicity consider
the principal
bundle G(G/H, H) first.
Because our vector bundle in question is the associated bundle
via the irreducible representation of H, the covariant derivative in
the
vector bundle will follow immediately from that of the principal
bundle.

Consider now the vector fields $X_m$ defined by the relation,
$$
X_m {\sigma}^{-1}(q)= {\sigma}^{-1}(q) T_m\,.
\eqn\killing
$$
These vector fields are just the generalizations
of the usual Killing vector fields regarded as first order
differential
operators.
(In the context of Berry's phase these are modified symmetry
generators for effective Hamiltonians [\Levay].)
The fact that such vector fields do exist
can be seen explicitly by examining the infinitesimal
version of $g\s(q) = \s(gq)h(g,q)$, which leads to
a first order differential operators for $X_m$ satisfying the
commutation
relations of the Lie algebra $\g$.

We shall then consider the following covariant derivative
$$
{\nabla}_m := - {\cal D}_m^n(\s^{-1}) X_n\ ,
\eqn\covderiv
$$
where ${\cal D}_m^n(\s)$ is the \lq adjoint matrix'
(the matrix of the adjoint
representation of G
in the basis $T_m$) defined by
$$
{\sigma}^{-1}(q) T_m {\sigma}(q) = {\cal D}_m^n(\s) T_n\ .
\eqn\conmatrix
$$
Using this, we may invert
(\killing)
to get
$$
{\cal D}_n^m(\s^{-1}) X_m \s^{-1} = T_n \s^{-1}\ .
\eqn\no
$$
We hence find that our covariant derivative (\covderiv) satisfies
$$
{\nabla}_m \s^{-1} = - T_m \s^{-1}\ ,
\eqn\simpleaction
$$
that is, it behaves just as $-T_m$ on $\s^{-1}$.

The $\r$ component of ${\nabla}_m$
is the covariant derivative with respect to the H-connection.
To see this, following
the standard line of argument [\Strathdee] one
decomposes the canonical 1-form as
${\sigma}^{-1}d{\sigma}=-d({\sigma}^{-1}){\sigma}= A^{\rm H} + e$
where $A^{\rm H} =  A^i_\a T_i dq^\a$ is the H-connection and
$e = \s^{-1} d\s\vert_{\r} = e_\a^{\,\,a} T_a dq^\a$
is the vielbein, with $T_i \in \h$ and $T_r \in \r$ in the
orthogonal decomposition $\g = \h \oplus \r$.
Using the inverse of the vielbein,
$e^\a_{\,\, a} e_\a^{\,\,b} = \delta_a^b$, one may cast the canonical
1-form into the vielbein frame.  This yields
$$
({\partial}_a + e^{\,\,\a}_a A_\a^i T_i) {\sigma}^{-1}
= - T_a {\sigma}^{-1} = {\nabla}_a{\sigma}^{-1}\ ,
\eqn\no
$$
where (\simpleaction) is used in the last equality, proving therefore
our claim.

When we go over to the vector bundle from the principal bundle, we
have to act with
the covariant derivative on the expansion (\harwave), hence we are to
use the particular representation $\chi$ for the generators $T_i$ of
$\h$.  It is then clear that
the covariant derivative acts in an extremely simple manner
on the wave functions.  In fact, the property (\simpleaction) shows
that
the covariant derivative in the representation $\chi$ is
indeed the representation of the element $T_m$
on such wave functions.  Hence, if we adopt
for the Hamiltonian the quadratic Casimir $ X_mX^m={\nabla}_m
{\nabla}^m$
of the group G --- which is G-invariant by construction --- we find
that the Hamiltonian is given by the square of the
covariant derivative ${\nabla}_a{\nabla}^a$ modulo a constant which is
the value of the quadratic Casimir of the subgroup H
evaluated on the irreducible representation
$\chi$.  Thus
the free,
homogeneous Hamiltonian
given by the quadratic Casimir
leads precisely to the Hamiltonian for the particle minimally
coupled to the H-connection,
that is, the Hamiltonian argued by Landsman and Linden [\LL].

\vfill\eject


\noindent \secno=3 \meqno=1

\noindent
{\bf 3. Quantizing on an $n$-Sphere}
\medskip
\noindent
In the approach to quantizating
on spheres $S^n$ proposed by Ohnuki and Kitakado [\OK]
there appeared (possibly topological)
gauge fields on the spheres as a result of
inequivalent quantizations.   These (infinitely) many
inequivalent quatizations are labelled by the
irreducible representations
of the group $SO(n)$ --- an important feature shared
with Mackey's approach [\Mackey] where one
regards $S^n$ as $SO(n+1)/SO(n)$.
Thus it would be natural to expect that the gauge fields
observed in [\OK] may coincide with the H-connection found
by Landsman and Linden [\LL] in Mackey's approach.
We shall show below that this is indeed the case.

But let us first recall the quantization and the
gauge fields discussed in [\OK].
There, quantization is prescribed by
embedding the sphere $S^n$ in
$\Real^{n+1}$
and then postulating a \lq fundamental algebra'
as a set of quantum relations,
generalizing the conventional canonical commutation
relations.  The fundamental algebra is the Lie algebra of
$E(n+1)$, the Euclidean group
in $n+1$ dimensions given by the semidirect product of
$SO(n+1)$ and $\Real^{n+1}$, and finding the Hilbert space
${\cal H}(S^n)$
amounts to finding the
representations of the group taking into account
the constraint that restricts to the sphere.
Wigner's technique then allows for constructing explicitly
the representations of $E(n+1)$
from the irreducible representations
of the subgroup $SO(n)$, which is the isometry
group of $SO(n+1)$ acting on $S^n$.
According to this, the representations
(of the Lie algebra) of $E(n+1)$
may be found by looking at the infinitesimal generators
of the Wigner rotation.  In Mackey's language
the Wigner rotation corresponds to the matrix element\note
{
We here assume for simplicity that $q$ and $g^{-1}q$ are in the
same patch where a single section $\s$ is available.
}
$$
Q_{\mu\nu} (g,q) :=
\pi^\chi_{\mu\nu}\bigl((\s(q))^{-1}g\s(g^{-1}q)\bigr)\ ,
\eqn\wr
$$
representing the rotations in the components of the vector-valued
wave function in the induced representation (\induced).
In the present case $g \in SO(n+1)$ and $q$ stands for
a vector on the sphere $S^n$ embedded in $\Real^{n+1}$, and we
take the radius of the sphere to be
unity, $\sum_{\a=1}^{n+1} (q^\a)^2 = 1$.
In this embedding we adopt the convention that any function
on $S^n$ is  smoothly extended to $\Real^{n+1}$ by
continuing the value of the function constantly along
the direction of the radius.  This implies that any function
$f(q)$ defined this way obeys the condition,
$q^\a \pa_\a f(q) = 0$, where $\pa_\a = \pa/ \pa q^\a$.

We label the basis of the Lie algebra $so(n+1)$ by antisymmetric
operators $T_{\a\b}$ with $\alpha$ and $\beta$ running over
$ 1, \ldots, n+1$.
The $so(n)$ subalgebra is identified with the generators $T_{ab}$,
where $a$ and $b$ can take values $ 1, \ldots, n$. The reductive
decomposition $so(n+1)=so(n)\oplus\r$ is then given by
$so(n+1)={\rm span}\,\{T_{ab}\}\oplus{\rm span}\,\{T_{a}\}$ where
$T_a=T_{a,n+1}$. The commutation relations are then
$$
\eqalign{
[T_{ab}, T_{cd}]
&= \d_{ad}T_{bc}+ \d_{bc}T_{ad}- \d_{ac}T_{bd}- \d_{bd}T_{ac}\ ,\cr
[T_{ab},T_c]&=\d_{bc}T_a-\d_{ac}T_b \ , \cr
[T_a,T_b]&=-T_{ab} \ .
}
\eqn\commutator
$$
To make the presentation easier we now
omit the label $\pi^\chi$ for the representation used.

Corresponding to the infinitesimal transformation
$
g = e^{{1\over2}\epsilon_{\alpha\beta}T_{\alpha\beta}}
     = 1 + {1\over2}\epsilon_{\alpha\beta}T_{\alpha\beta}
$
with $\epsilon_{\alpha\beta}$ being real antisymmetric parameters,
we have the Wigner rotation,
$$
Q (g,q) = 1 + \tfrac12\epsilon_{\alpha\beta} f_{\alpha\beta}(q)\ ,
\eqn\wrinf
$$
where $f_{\alpha\beta}(q)$ are the generators of the rotation.
Then, the combination [\OK]
$$
A_\alpha(q) := f_{\alpha\beta}(q)\, q^\beta\ ,
\eqn\pot
$$
is seen to appear in the Hamiltonian in the form
covariantly coupled to
a particle, and hence is regarded as an
induced gauge field.
We now show
that this gauge field (\pot) is in fact the H-connection.

To this end, observe first that from (\wr)
the generators in (\wrinf) are given by
$$
f_{\alpha\beta}(q) = \s^{-1}(q) T_{\alpha\beta} \s(q)
 - \s^{-1}(q) \pa_\mu \s(q)
 {{\pa q^\mu(\epsilon)}\over{\pa
\epsilon_{\a\b}}}\big\vert_{\epsilon=0}\ ,
\eqn\wrinfb
$$
where
$q^\mu(\epsilon) := (g\,q)^\mu=q^\mu+\tfrac12
\epsilon_{\a\b}(T^{\rm def}_{\a\b})_{\mu\nu}q^\nu$,
and $(T^{\rm def}_{\a\b})_{\mu\nu}
=\d_{\a\mu}\d_{\a\nu}-\d_{\a\nu}\d_{\b\mu}$ is
the defining representation of $so(n+1)$. From this we get
$$
\frac{\pa q^\mu(\epsilon)}{\pa \epsilon_{\a\b}}\big\vert_{\epsilon=0}
=\d^{\a\mu}q^\b-\d^{\b\mu}q^\a\,.
\eqn\no
$$
It is then easy to see that
under the change of section $\s(q) \rightarrow
\s(q) h(q)$ for some $h(q) \in SO(n)$ the gauge
field (\pot) transforms as a connection,
$$
A_\a(q) \rightarrow
h^{-1}(q) A_\a(q) h(q) - h^{-1}(q)\pa_\a h(q)\ .
\eqn\trsftop
$$
This is also evident from the expression,
$$
A_\a(q) = \s^{-1}(q) T_{\alpha\beta}q^\b \s(q)
 - \s^{-1}(q) \pa_\a \s(q)\ ,
\eqn\ok
$$
obtained from the definition (\pot).

Consider now the section
$$
\s(q) = e^{\theta^a(q) T_a}\,,
\eqn\section
$$
which provides a local mapping from $S^n$ to ${\rm G}=SO(n+1)$.
The inverse mapping is given by
$$
q^a := \theta^a {{\sin \vert\theta\vert}\over{\vert\theta\vert}},
\quad a=1, \ldots, n, \qquad
q^{n+1} := \cos \vert\theta\vert\, ,
\eqn\cd
$$
where
$\theta^t = (\theta^1, \ldots, \theta^n)$ and $\vert\theta\vert =
\sqrt{\theta^t \theta} = \sqrt {\sum_{a} (\theta^a)^2}$.
With the section (\section) one finds that the relevant parts of
the adjoint matrix (\conmatrix),
$$
\eqalign{
\s^{-1}(q)T_{ab}\s(q)&={\cal D}^{cd}_{ab}\,
T_{cd}+{\cal D}^{c}_{ab}\,T_c\ ,\cr
\s^{-1}(q)T_{a}\s(q)&={\cal D}^{bc}_{a}\,T_{bc}+
{\cal D}^{b}_{a}\,T_b\ ,\cr
}
\eqn\autobits
$$
take the form [\Levay]
$$
{\cal D}^{bc}_a=\tfrac12(q^b\d^c_a-q^c\d^b_a)\,,\eqn\autoone
$$
and
$$
{\cal D}^{cd}_{ab}=\tfrac12(\d^c_a\d^d_b-\d^c_b\d^d_a)
+\frac{q_b(q^c\d^d_a-q^d\d^c_a)+q_a(q^d\d^c_b-q^c\d^d_b)}
{2(1+q^{n+1})}\,.
\eqn\autotwo
$$

To show that (\ok) is the H-connection we note that
the $\h$-part in the first term on the right hand side of (\ok) vanishes,
$$
\s^{-1}(q)\, T_{\alpha\beta}q^\b \s(q)\vert_\h = 0\ .
\eqn\noh
$$
For $\a = n+1$, this is obvious since the middle piece
$T_{\alpha\beta}q^\b$ that
is conjugated under $\s(q)$ is precisely
proportional to the argument in the exponential of $\s(q)$;
see (\section) and (\cd).
For $\a=a\ne n+1$, using (\autoone), (\autotwo) and the antisymmetry
of
$T_{cd}$, we have
$$
\s^{-1}(q)T_{a\b}q^\b\s(q)\vert_\h=(q^b{\cal D}_{ab}^{cd}+q^{n+1}
{\cal D}_a^{cd})T_{cd}=0\,,\eqn\no
$$
which establishes (\noh).

Now since the gauge field (\ok) must
lie anyway in the space $\h = so(n)$ by construction (because
it is formed out of the generators of the $SO(n)$ Wigner rotation),
we see that the $\r$-part
of the two terms in the right hand side of
(\ok) must precisely cancel each
other.
Combined with (\noh), this implies that
$$
A_\a(q) = - \s^{-1}(q) \pa_\a \s(q)\vert_\h\ ,
\eqn\aim
$$
that is, Ohnuki-Kitakato's gauge field (\pot) is in fact the
H-connection (up to the irrelevant sign).
In terms of the section (\section)
the H-connection reads\note
{The confirmation by a direct computation is also given
in [\Tanimura].}
$$
 \s^{-1}(q) d \s(q)\vert_\h
= {1 \over {1 + q^{n+1}}} \sum_{a,b}^n q_a d q_b \, T_{ab}\ ,
\eqn\no
$$
which of course agrees with the expression found in [\OK].

In passing, we mention that in a recent paper [\FKO] it is pointed out
that the gauge field (\pot) can be mapped into
the \lq generalized BPST instanton' solution found earlier [\Fujii]
--- a solution
of the Yang-Mills equation on $S^n$ which is topologically
nontrivial for $n$ even and trivial for $n$ odd.  The above result
implies that this solution is essentially identical to the
H-connection,
although the meaning of self-duality can change under the mapping.

\vfill\eject

\noindent \secno=4 \meqno=1

\noindent
{\bf 4. Berry's connection as the H-connection}
\medskip
\noindent
Berry's phase arises in systems where the Hamiltonian has degenerate
eigenstates
labelled by a collection of parameters, which are identified with the
slow degrees of freedom. Adiabatically decoupling the fast variables
from these slow ones results in an effective theory with a gauge
structure
in the slowly varying system [\Shapere].
The form of the gauge field that emerges is
governed by the geometry of the slow system.
In applications the
degeneracies reflect a symmetry of the system, hence the slow
system is usually identified with a coset space G/H. Such an
identification emerges from a Hamiltonian of the form
$$
H(q)=U(q) H_0 U^{-1}(q)\,,\eqn\berryham
$$
where $q\in{\rm G/H}$ are the slow variables, $U(g)$ is a unitary
irreducible representation of G and $H_0$ is typically
an element of the
enveloping algebra of the subgroup H,
commuting with the restriction of the
representation $U$ to H.
It is readily confirmed [\Vinet, \Giler, \Levay] that if
we let $U(q)$ be in the form $U(\s(q))$, then (\sol)
furnishes the eigenstates of the
Hamiltonian with $\ket{\chi, \bar\mu}$ being the eigenstates of
$H_0$ labelled by some irreducible representation $\chi$ of H.
Thus our representation $\chi$ is obtained
from the given representation of G
by restriction to H, followed by a further restriction
to an invariant subspace.
Using the states (\sol) (again dropping the label
$\a$ for the patch to which $q$ belongs) Berry's connection reads
$$
\eqalign{
\sum_m A^{\rm Berry}_m(q) (T_m)_{\mu\nu}&=\bra{q,\chi,\mu}
                                           d\ket{q,\chi,\nu}\cr
&= \bra{\chi,\bar{\mu}} \, U^{-1}\bigl(\s(q)\bigr)\,
dU\bigl(\s(q)\bigr)\,\ket{\chi,\bar{\nu}}\cr
&=\sum_i A^{\rm H}_i(q) (T_i)_{\mu\nu}+\sum_ae_a(q)
\bra{\chi,\bar\mu}U(T_a)
\ket{\chi,\bar\nu}
}\eqn\berrycon
$$
The identification of
this connection with  the H-connection clearly depends
on whether the final term is zero or not. In applications
this term is often set equal to zero by hand [\Giler, \Levay].
That this term is not always
zero, though, is best seen through an explicit example.

Consider the situation where the slow variables parametrize a three
sphere $S^3$, now viewed as the coset space $SO(4)/SO(3)$.
This would arise, for example,
from (\berryham) by taking $H_0$ to be the quadratic Casimir for
$SO(3)$.
In the Lie algebra of $SO(4)$ we take the reductive decomposition
$$
\g = \h\oplus\r={\rm span}\{ T_i \} \oplus {\rm span} \{ T_a \}\,,
\qquad i, \, a = 1, \,2, \,3\,.
\eqn\no
$$
with the $T_i$'s forming an $su(2)$ algebra, $[T_i,T_j]=\e_{ijk}T_k$,
and
the remaining commutators being
$$
[T_i,T_a]=\e_{iab}T_b\ , \eqn\vector
$$
and
$$
[T_a,T_b]=\e_{abi}T_i\,.\eqn\shift
$$
The non H-connection part of Berry's connection is, in this example,
$\sum_a e_a\bra{jm}T_a\ket{jm'}$, where we have reverted to the
familiar
notation for the representation of angular momentum. We now
show that the matrix element  $\bra{jm}T_a\ket{jm'}$ need not be zero
in general.

For this, we
note first
that the commutator (\vector)
implies that the basis vectors in $\r$ transform
as a vector (spin 1)
operator. To emphasise this fact we will, henceforth,
denote these operators by $T^{(1)}_a$.
The Wigner-Eckart theorem then tells us that the $m$, $m'$ and $a$
dependence of this matrix element resides in the Clebsch-Gordan
coefficients $\bra{jm'1a}jm\>$:
$$
\bra{jm}T^{(1)}_a\ket{jm'} =
\bra{jm'1a}jm\> \<j\|T^{(1)}\|j\>\ ,
\eqn\wet
$$
where $\<j\|T^{(1)}\|j\>$ is the reduced matrix element
which is independent of $m$, $m'$ and $a$.
In terms of the basis $T_\pm := i (T_1 \pm iT_2)$, $T_0 := iT_3$,
the Clebsch-Gordan
coefficients are given by
$$
\bra{jm'1a}jm\>={{\delta}_{m'+a,m}\over
\sqrt {j(j+1)}} \cases{\mp\sqrt{(j\pm m)(j \mp m+1)/2},&if $a =\pm
$;\cr
m,& if $a =0$.\cr}
\eqn\cg
$$
Upon identifying the same $a$ and $i$, we find that
these coefficients are related to the
representation matrix elements
$\bra{jm'}T_i\ket{jm}$
of the $su(2)$ generators $T_i$, $i= +,-,0$.  This allows us to
rewrite (\wet) as
$$
\bra{jm}T_a^{(1)}\ket{jm'}=a_j\bra{jm'}T_i\ket{jm}\ ,  \eqn\no
$$
where the prefactor $a_j$ is
$$
a_j=-\frac{\<j\|T^{(1)}\|j\>}{\sqrt{j(j+1)}}\,.\eqn\aj
$$
We recall that the action of any vector operator on the state
$\ket{jm'}$
is determined by two reduced matrix elements. For $T^{(1)}$ these are
$a_j$ and the reduced matrix element $\<j-1\|T^{(1)}\|j\>$.
However, the action of $T^{(1)}$ is also fixed by the fact that it
comes from a representation
of $SO(4)$. Exploiting these two facts allows us to determine the
allowed values for $a_j$.

The irreducible unitary representations of $SO(4)$ are labelled
by two numbers $(k_0,c)$, where $k_0=0,\frac12,1,\frac32,\dots$ and
$c=\pm(k_1+1)$ with
$k_1=k_0,k_0+1,k_0+2,\dots$. (For a clear account of this see
[\Bohm].)
The representation space is then decomposed into the direct sum
$$
{\cal R}(k_0,c)=\bigoplus_{j=k_0}^{k_1}{\cal R}^j\,,\eqn\no
$$
of the irreducible representations ${\cal R}^j$
of $SO(3)$ spanned by the angular momentum states
$\ket{j m}$, $m=-j,\dots, j$.
In such a representation one finds that $a_j$ is given by
$$
a_j=\frac{k_0c}{j(j+1)}\,.\eqn\ajis
$$
{}From this we deduce that if $k_0\ne0$ then Berry's connection does
not
correspond to the H-connection.

This example can be extended to more general coset spaces in much the
same way by using the generalized Wigner-Eckart theorem (see, for
example,
[\Wudka]). The conclusion
reached is that, in general, Berry's connection is not the
H-connection.
The question we now want to address is what additional structures
are needed in order to ensure that they do coincide.
To motivate our analysis
of this problem it is again
useful to return to the three sphere example discussed above.

{}From (\ajis) we see that the relevant reduced matrix element
vanishes
only when $k_0=0$. In this case (and only in this case) the
representations
$(k_0=0,c)$ and $(k_0=0,-c)$ of $SO(4)$ are unitarily equivalent
(there is no parity
doubling [\Bohm]). The representation space becomes the direct sum
$$
{\cal R}(0,n)=\bigoplus_{j=0}^{n-1}{\cal R}^j \ ,
\qquad\hbox{where}\quad n=1,2,\dots.
\eqn\no
$$
The action of $T_i$ on ${\cal R}^j$ is the standard one, changing the
value of $m$ by $\pm1$. From (\vector) one can also
show that the action of $T^{(1)}_a$ on
${\cal R}^j$ changes the value of $j$ by $\pm1$. Thus the state
$\ket{j m} =
\ket{n-1,n-1}$ is both a highest weight vector for the irreducible
representation on ${\cal R}^{n-1}$ of $SO(3)$, and for the irreducible
representation on ${\cal R}(0,n)$ of $SO(4)$. This cannot hold for any
of the other ($k_0\ne0$) representations of $SO(4)$ since the parity
doubling found in those representations would then imply that such
a vector was a highest weight for two inequivalent representations.

We shall use this example as a motivation for the following restriction
on the allowed states $\ket{\chi,\bar\mu}$ that occur in (\berrycon).
Recall first that by definition the reference basis states satisfy
$$
U(h)\vert\chi,\bar{\mu}\rangle =
\sum_\nu \vert\chi,\bar{\nu}\rangle{\pi}_{\nu\mu}^{\chi}(h),\qquad
\hbox{for} \quad h \in {\rm H}\,.
\eqn\repres
$$
Let $\Lambda$ be the highest weight
labelling the representation of the group G
in question.  We shall
consider the {\it highest subspace} ${\cal H}_{\Lambda}$, which
is the
subspace of the representation space ${\cal H}$ of G realizing
(\repres)
and also contains the vector $\vert\Lambda\rangle$ corresponding to
the highest weight.
We then claim that, for a wide class of systems,  by choosing the
subspace
as a highest subspace we will manage to obtain
merely the $\h$-part of Berry's connection.
To prove this it is convenient to
develop an alternative description of the highest subspace
${\cal H}_\Lambda$.

For this, let us restrict ourselves to cosets G/H,
where the subgroup H is given by the centralizer ${\rm S}_K$
of some element $K \in \g$.
This corresponds
to the Hamiltonian (\berryham) whose parameter space
is the coadjoint orbit of the group G passing through $K$ discussed
in [\Vinet, \Levay].  If $K$ is a regular semisimple element [\Hum]
of $\g$
then H in this case is just the Cartan subgroup T regarded
as the maximal torus containing $K$, but if not
then H is greater than T.
Let $\Sigma$ be the root system of G relative to T, and let
${\Sigma}_K$
be the root system of H relative to T. By considering
the
complexification $\g_{\rm c}$ of ${\g}$  we have the Cartan
decomposition
$$
\g_{\rm c}=\t_{\rm c}\oplus {\sum}_{\a\in\Sigma}
\g_\a\ ,
\eqn\cartanrootg
$$
where $\t_{\rm c}$ is the complexification of the Cartan subalgebra
$\t$,
and $\g_\a$ is the root space corresponding to the root
$\a$.
Similarly we have
$$
\h_{\rm c}=\t_{\rm c}\oplus {\sum}_{\a\in{\Sigma}_K}
\h_{\alpha}\ .
\eqn\cartanrooth
$$
Next, let $W$ be a Weyl chamber of $\t$ relative to G, and $W_K$ be
a Weyl
chamber of $\t$ relative to H.  We can
define
the positive roots ${\Sigma}^{+}$ (${\Sigma}_K^+$) of ${\Sigma}$
(${\Sigma}_K$)
with respect to $W$ ($W_K$).
It is then guaranteed [\Wallach] that there exists a
\lq $K$ admissible Weyl chamber\rq\
satisfying: \break (i)
${\Sigma}^+\cap{\Sigma}_K={\Sigma}_K^+$, and (ii)
if $\a\in{\Sigma}^+ -{\Sigma}_K^+$,
$\b\in{\Sigma}_K$ and $\a +\b \in {\Sigma}$, then $\a +\b \in
{\Sigma}-{\Sigma}_K^+$.

Armed with this, we then show
that the highest subspace ${\cal H}_\Lambda$ can alternatively
be characterized by
$$
{\cal H}_{\Lambda}=\bigl\{\, \vert\phi\rangle\in {\cal H}
\, \big\vert \,\,\,
U(T_{\alpha})\vert\phi\rangle =0,
\quad \forall\alpha\in{\Sigma}^+
-{\Sigma}_K^+\bigr\}\ .
\eqn\otherdescription
$$
Note first that the states defined by
(\otherdescription) are invariant under the action of H in ${\cal H}$.
Indeed,
for those generators of $\h$ belonging to the Cartan subalgebra $\t$
this
is obvious since for $T_i\in\t$, $[T_i,T_\a]=\a(T_i)T_\a$. If $T_\b$
is a
generator of $\h$ not in $\t$ then, using the reductivity of the
decomposition $\g=\h\oplus\r$,
we get, for $\a\in\Sigma^{+}-\Sigma^{+}_K$,
$$
U(T_{\alpha})U
(T_{\beta})\vert\phi\rangle =
U(T_{\beta})U(T_{\alpha})\vert\phi\rangle
+U([T_{\alpha},T_{\beta}])\vert\phi\rangle
=C_{\alpha\beta}^{\alpha +\beta}U(T_{\alpha +\beta})
\vert\phi\rangle\ ,
\eqn\no
$$
which vanishes
since ${\alpha +\beta}\in {\Sigma}^+ -{\Sigma}_K^+$.

Second, the unitary action in (\otherdescription)
is also irreducible.  To see
this, suppose that it is reducible. Then there exists some
$\vert\Omega\rangle\neq
\vert\Lambda\rangle$ for which
$$
U(T_{\beta})\vert\Omega\rangle =0,
\qquad \hbox{for} \quad \beta\in{\Sigma}_K^+\ ,
\eqn\no
$$
{\it i.e.,} $U(T_{\beta})$ is a step operator in ${\h}$
annihilating this
state. It then follows that both  the operators
$U(T_{\alpha})$ and
$U(T_{\beta})$, where $\alpha\in{\Sigma}^+ -{\Sigma}_K^+$ and
$\beta\in {\Sigma}_K^+$, annihilate $\vert\Omega\rangle$.
Hence this state is annihilated by any $U(T_{\alpha})$
for $\alpha\in {\Sigma}^+$, which implies that $\vert\Omega\rangle$
is a highest weight.
But since we cannot have two highest weights,
we see that ${\cal H}_{\Lambda}$ defined by (\otherdescription) is
irreducible and hence must be the
highest subspace satisfying (\repres).

Having established (\otherdescription),
we now find, for such highest subspace states,
$$
\sum_{\a\in\Sigma-\Sigma_K}
e_\a(q)\bra{\chi,\bar\mu}U(T_\a)
\ket{\chi,\bar\nu} = 0\ ,
\eqn\no
$$
on account of $T_{-\alpha}=
{T_{\alpha}}^{\dagger}$.
Clearly, then, we can conclude that
if a highest subspace is used in the construction of
Berry's connection, then there will be no
$ \r$-part and hence it will be the H-connection --- the
claim we wished to prove.

\vfill\eject


\noindent \secno=5 \meqno=1

\noindent
{\bf 5. Conclusions and Discussions}
\medskip
\noindent
In this paper we have argued that the induced connection that appears
on a coset space
in Mackey's quantization scheme admits a natural interpretation, that is,
it arises from the homogeniety criterion
required for the Hamiltonian.  This led to an alternative account from
[\LL]
of why the Hamiltonian
on G/H involves the induced H-connection.
Being geometrical, our criterion
will be useful even in other quantization approaches and, possibly,
in attempts at extending the quantization scheme to field theories.
Indeed, we have shown that the gauge field induced in a slightly
different approach [\OK] is again the H-connection --- a fact
suggesting
a universal feature of the quantum theory on such topologically
non-trivial
spaces.  In connection with this, it is worth mentioning that
even in the \lq confining approach' [\JK] to quantization,
which is totally different from Mackey's approach, one can still
observe
an induced gauge field
which also appears to be of the type of the H-connection
[\TT, \FO].

The appearance of the H-connection
in the other context --- Berry's phase --- was then analyzed in the
setting
where the parameter space is given by a coset space G/H.  We have seen
that Berry's connection becomes the H-connection if the energy
eigenspace
we are looking at possesses the highest weight state of the unitary
representation of the group G that characterizes the system.
Notice also that such highest subspaces can be used to define the so
called vectorial coherent states [\ZG] for the group G.
Indeed, by choosing the states $\ket{\chi, \bar\mu}$ as the ones
belonging
to a highest subspace, the states of (\sol) become the
vectorial coherent states.
The physical implications of this condition for the  energy
eigenspaces need to be investigated, but
we have at least seen an interesting fact that
in such cases
the effective theory describing the slow variables
bears an unexpected resemblance with the quantum theory on coset
spaces.

\vskip 1cm
\noindent{\bf Acknowledgement}

\noindent
We wish to thank Y.S. Wu for useful discussions and encouragement.

  \vfill\eject\immediate\closeout\reffile
  \centerline{{\bf References}}\bigskip\eightpoint\frenchspacing%
  \input refs.tmp\vfill\eject\nonfrenchspacing


\bye